\documentclass[aps,prd,preprintnumbers,showpacs,showkeys,nofootinbib,superscriptaddress,fleqn,floatfix,tightenlines,10pt]{revtex4-2} 
\usepackage{amsmath,amssymb}
\usepackage{graphicx,bm}
\usepackage{slashed,color}
\usepackage[colorlinks=true,pdfstartview=FitV,bookmarks=true,bookmarksnumbered=true, 
breaklinks]{hyperref}

\hypersetup{linkcolor=red, citecolor=blue, urlcolor=blue}

\usepackage[normalem]{ulem} 
\usepackage[dvipsnames]{xcolor} 
\renewcommand\sout{\bgroup \color{red} \ULdepth=-.5ex \ULset}

\begin{document}
\preprint{INHA-NTG-05/2026}
\title{Nonexistence of hidden-charm pentaquarks in $J/\psi$
  photoproduction} 
\author{Samson Clymton}
\email{samson.clymton@apctp.org}
\affiliation{Asia Pacific Center for Theoretical Physics (APCTP),
  Pohang, Gyeongbuk 37673, South Korea}
\author{Sang-Ho Kim}
\email{shkimphy@gmail.com}
\affiliation{Department of Physics and Origin of Matter and Evolution
  of Galaxies (OMEG) Institute, Soongsil University, Seoul 06978,
  South Korea} 
\author{Hyun-Chul Kim}
\email{hchkim@inha.ac.kr}
\affiliation{Department of Physics and Institute of Quantum Science,
  Inha University, Incheon 22212, South Korea}
\affiliation{School of Physics, Korea Institute for Advanced Study
  (KIAS), Seoul 02455, Republic of Korea} 

\date{\today}

\begin{abstract}
We investigate $J/\psi$ photoproduction off the proton, $\gamma p \to
J/\psi p$, to elucidate the nonexistence of hidden-charm pentaquark
signals reported by the GlueX and CLAS12 experiments. Within a
coupled-channel rescattering mechanism, we employ the transition
amplitudes from a previous coupled-channel analysis that dynamically
generates the $P_{c\bar{c}}$ states. The kernel amplitudes for the
transition to the $J/\psi N$ channel include both $t$-channel
heavy-meson exchange and $u$-channel heavy-baryon exchange. We find
that the rescattering contributions from the $\bar{D}^{(*)}\Sigma_c$
channels---indispensable for the formation of the $P_{c\bar{c}}$
resonances---are about one order of magnitude smaller than those from
$\bar{D}^{(*)}\Lambda_c$, since $g_{\bar{D}^{(*)}N\Sigma_c}$ is
roughly five times smaller than $g_{\bar{D}^{(*)}N\Lambda_c}$. Since
the $P_{c\bar{c}}$ resonances couple to the $J/\psi N$ channel
predominantly through the $\bar{D}^{(*)}\Sigma_c$ intermediate states,
their suppression prevents the pentaquark signal from appearing in
photoproduction. With only a single parameter controlling the overall
normalization, the present work describes the GlueX and CLAS12 cross
sections well. These results suggest that the null result from
photoproduction need not be in conflict with the pentaquark signals
observed by the LHCb Collaboration.   
\end{abstract}

\maketitle

\section{Introduction}
Since the LHCb Collaboration announced the existence of hidden-charm
pentaquark baryons, a plethora of experimental and theoretical works
have been performed. These states were first observed in the
$J/\psi p$ invariant mass spectrum from $\Lambda_b^0 \to J/\psi p
K^{-}$ decays~\cite{LHCb:2015yax,LHCb:2019kea}. Thus far, four of them
have been established: three narrow states below the
$\bar{D}^* \Sigma_c$ threshold and one broad state below the
$\bar{D}\Sigma_c^*$ threshold. Notably, $P_{c\bar{c}}(4330)$ was
observed in $B_s^0\to J/\psi p\bar{p}$ decays, with no signal from the
previously reported $P_{c\bar{c}}(4312)$~\cite{LHCb:2021chn}. In
contrast, the GlueX~\cite{GlueX:2019mkq,GlueX:2023pev} and CLAS12~\cite{CLAS:2026lls} Collaboration found no evidence for such states in
$J/\psi$ photoproduction off the
nucleon. Given that the
$P_{c\bar{c}}$ states decay almost exclusively into $J/\psi$ and a
proton, this null result is rather surprising. Several works
subsequently tried to explain this
absence~\cite{Du:2020bqj, JPAC:2023qgg, Zhang:2024dkm,
  Sakinah:2026bhv}, but they mainly reproduced the experimental data
without clarifying the underlying mechanism.

We start from a recent coupled-channel formalism for hidden-charm
$J/\psi$ and open-charm two-body processes~\cite{Clymton:2024fbf},
which provides fully off-shell transition amplitudes. Based on it, we
show why hidden-charm pentaquarks cannot be generated in $J/\psi$
photoproduction. There, seven peaks were found, six identified as
$P_{c\bar{c}}$ resonances and one as a cusp. Four reproduce the known
states $P_{c\bar{c}}(4312)$, $P_{c\bar{c}}(4380)$,
$P_{c\bar{c}}(4440)$, and $P_{c\bar{c}}(4457)$, whereas two near
$4.5$~GeV, $P_{c\bar{c}}(4517,\,J^P=3/2^-)$ and
$P_{c\bar{c}}(4522,\,J^P=5/2^-)$, are predicted as
$\bar{D}^*\Sigma_c^*$ molecular states. Two further $J^P=1/2^+$
resonances, at $(4401-i35)$ and $(4533-i17)$~MeV, lack a clean
molecular interpretation and may correspond to genuine pentaquark
configurations. A dynamical explanation for the missing signal was
also proposed there through the $J/\psi N$ elastic channel, where the
negative-parity peaks appear as dips and are washed out by the larger
positive-parity ($P$-wave) bumps, while the resonances remain visible
in the $\Lambda_b \to J/\psi p K^-$-type transitions seen by LHCb.

In this work, we aim at clarifying the theoretical origin of this
suppression. To implement the transition amplitudes for $J/\psi$
photoproduction, the $u$-channel diagrams must be taken into
account. The crucial point is that the coupling constant at the
$\bar{D}\Sigma_c N$ vertex is much smaller than that at the
$\bar{D}\Lambda_c N$ one. In Ref.~\cite{Clymton:2024fbf}, the
$\bar{D}\Sigma_c$ channel was shown to be essential for describing the
resonances, in particular $P_{c\bar{c}}(4312)$. In $J/\psi$
photoproduction, however, the smallness of the $\bar{D}\Sigma_c N$
coupling constant prevents the dynamical generation of the
hidden-charm pentaquarks, which we demonstrate explicitly below.

The present paper is organized as follows. In Sec.~\ref{sec:2}, we
describe the general formalism for $J/\psi$ photoproduction. We first
construct the $\gamma N\to J/\psi N$ amplitudes from the effective
Lagrangian and then combine them with the two-body hidden-charm and
open-charm processes. Solving the coupled-channel
Blankenbecler-Sugar (BbS)-type rescattering equations, we obtain the
numerical results. In Sec.~\ref{sec:3}, we first present and compare
the results for $\bar{D}^{(*)}\Lambda_c \to J/\psi N$ and
$\bar{D}^{(*)}\Sigma_c \to J/\psi N$, where $\bar{D}^{(*)}$ generically
denotes either $\bar{D}$ or $\bar{D}^*$. We then display the total
cross section for $J/\psi$ photoproduction off the proton, compare it
with the GlueX and CLAS12 data, and show explicitly how the hidden-charm
pentaquarks fail to emerge. The last section is devoted to a summary
and conclusions.

\section{Formalism}
\label{sec:2}
\begin{figure}[htp]
\centering
\includegraphics[scale=0.45]{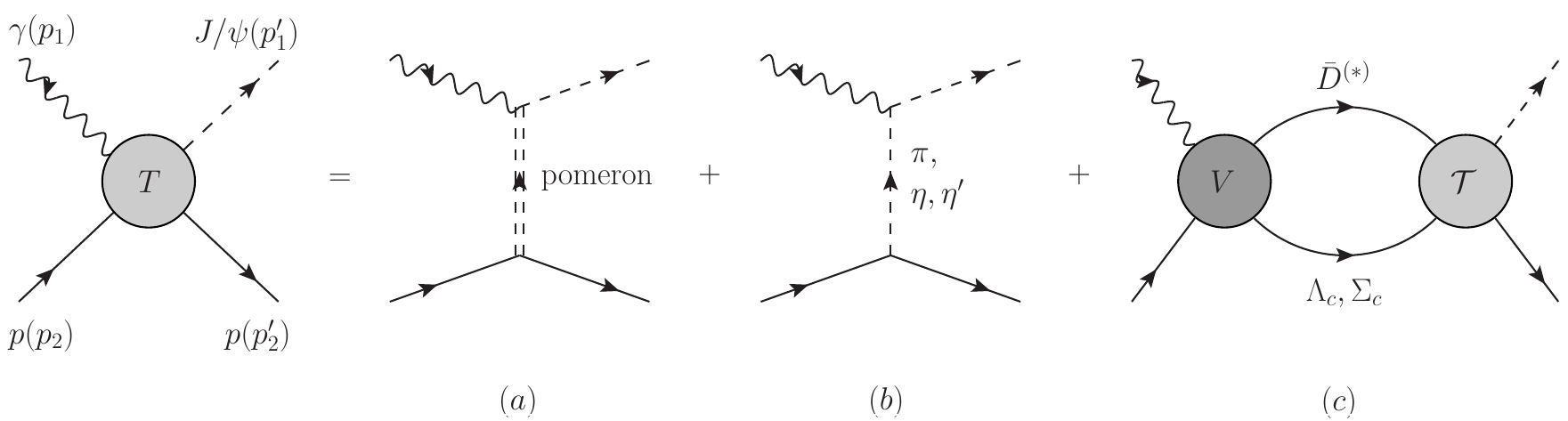}
\caption{Diagrams for $J/\psi$ photoproduction off the proton ($\gamma
  p \to J/\psi p$): (a) pomeron exchange, (b) the Born diagrams, and
  (c) the rescattering diagram.}
\label{fig:1}
\end{figure}

In this section, we explain the general formalism employed in the
present work. Taking into account the rescattering effects that
contain information on the dynamical generation of the hidden-charm
pentaquark states (see Fig.~\ref{fig:1}), we derive the transition
amplitudes for $J/\psi$ photoproduction by solving the following
coupled-channel BS-type rescattering equation:
\begin{align}
  T_{J/\psi p,\gamma p} (\bm{p}',\bm{p}) =\, V_{J/\psi p,\gamma p}
  (\bm{p}',\bm{p})
  +\frac{1}{(2\pi)^3}\sum_k\int
  \frac{d^3q}{2E_{k1}(\bm{q})\,E_{k2}(\bm{q})}\,
  V_{k,\gamma
  p}(\bm{q},\bm{p})\,\frac{E_k(\bm{q})}{s-E_k^2(\bm{q})+i\varepsilon}\,
  \mathcal{T}_{J/\psi p,k}(\bm{p}',\bm{q}),
\label{eq:1}
\end{align}
where $V_{k,\gamma p}$ denotes the two-body Feynman kernel amplitude
for the intermediate state $k$ in the rescattering equation. When
$k = J/\psi\,p$, $V_{J/\psi p,\gamma p}$ represents the
$\gamma p\to J/\psi p$ Born diagrams~\cite{Kim:2025oyo}. 
$\mathcal{T}_{J/\psi p,k}$ stands for the transition
amplitude for the $J/\psi p \to k$ process, where $k$ runs over the
intermediate two-body open-charm states: $\bar{D}\Lambda_{c}$,
$\bar{D}^{*}\Lambda_{c}$, $\bar{D}\Sigma_{c}$,
$\bar{D}\Sigma_{c}^{*}$, $\bar{D}^{*}\Sigma_{c}$, and
$\bar{D}^{*}\Sigma_{c}^{*}$. Since the elastic amplitude
$\mathcal{T}_{J/\psi p,J/\psi p}$ is much smaller than those involving
open-charm intermediate channels, it is excluded in the present work.
The $\mathcal{T}_{J/\psi p,k}$ amplitudes were already constructed in
Ref.~\cite{Clymton:2024fbf}. Here, $\bm{p}$ and $\bm{p}'$ are the
relative three-momenta of the initial and final states, respectively,
while $\bm{q}$ denotes the three-momentum of the intermediate state in
the center-of-mass (CM) frame. The variable $s$ is the square of the
total CM energy, and $E_k = E_{k1}+E_{k2}$ is the total on-mass-shell
energy of the intermediate state.

We perform the partial-wave expansion of Eq.~\eqref{eq:1} in the
helicity basis and obtain the partial-wave transition amplitudes as
follows:
\begin{align}
T^{J}_{\lambda'\lambda} (\mathrm{p}',\mathrm{p}) =
V^{J}_{\lambda'\lambda} (\mathrm{p}',\mathrm{p})+
\frac{1}{(2\pi)^3} \sum_{k,\lambda_k} \int
\frac{\mathrm{q}^2\,d\mathrm{q}}{2E_{k1}E_{k2}}\,
V^{J}_{\lambda_k\lambda} (\mathrm{q},\mathrm{p})\,
\frac{E_k}{s-E_k^2+i\varepsilon}\,
\mathcal{T}^{J}_{\lambda'\lambda_k}(\mathrm{p}',\mathrm{q}),
\label{eq:2}
\end{align}
where the helicities of the final, initial, and intermediate states
are denoted by $\lambda'=\{\lambda'_1,\lambda'_2\}$,
$\lambda=\{\lambda_1,\lambda_2\}$, and
$\lambda_k=\{\lambda_{k1},\lambda_{k2}\}$, respectively. The channel
indices are dropped to simplify the notation. The variables
$\mathrm{p}'$, $\mathrm{p}$, and $\mathrm{q}$ represent the magnitudes
of the corresponding three-momenta $\bm{p}'$, $\bm{p}$, and $\bm{q}$,
respectively. The partial-wave expansion of the kernel amplitude
$V^{J}_{\lambda'\lambda}$ is given by
\begin{equation}
V^{J}_{\lambda'\lambda}(\mathrm{p}',\mathrm{p}) =
2\pi \int d(\cos\theta)\,
d^{J}_{\lambda_1-\lambda_2,\,\lambda'_1-\lambda'_2}(\theta)\,
V_{\lambda'\lambda}(\mathrm{p}',\mathrm{p},\theta),
\label{eq:3}
\end{equation}
where $\theta$ is the scattering angle and
$d^{J}_{\lambda\lambda'}(\theta)$ denotes the reduced Wigner
$D$-functions. Since the propagator in Eq.~\eqref{eq:2} develops a
singularity when the off-shell energy coincides with the on-shell one,
we regularize the rescattering equation by isolating the singularity
and treating it separately. Accordingly, the rescattering equation is
decomposed into regularized and singular parts as follows:
\begin{align}
  T^{J}_{\lambda'\lambda} (\mathrm{p}',\mathrm{p}) =
  V^{J}_{\lambda'\lambda} (\mathrm{p}',\mathrm{p})
  + \frac{1}{(2\pi)^3}
  \sum_{k,\lambda_k}\left[\int_0^{\infty}d\mathrm{q}\,
  \frac{\mathrm{q}\,E_k}{E_{k1}E_{k2}}
  \frac{\mathcal{F}(\mathrm{q})-\mathcal{F}(\tilde{\mathrm{q}}_k)}{s-E_k^2}
  + \frac{1}{2\sqrt{s}}
  \left(\ln\left|\frac{\sqrt{s}-E_k^{\mathrm{thr}}}{\sqrt{s}
  +E_k^{\mathrm{thr}}}\right|-i\pi\right)
  \mathcal{F}(\tilde{\mathrm{q}}_k)\right],
  \label{eq:4}
\end{align}
where
\begin{align}
  \mathcal{F}(\mathrm{q})=\frac{1}{2}\mathrm{q}\,
  V^{J}_{\lambda_k\lambda}(\mathrm{q},\mathrm{p})\,
  \mathcal{T}^{J}_{\lambda'\lambda_k}(\mathrm{p}',\mathrm{q}).
  \label{eq:5}
\end{align}
Here, $\tilde{\mathrm{q}}_k$ denotes the on-shell momentum defined by
the condition
$E_{k1}(\tilde{\mathrm{q}}_k)+E_{k2}(\tilde{\mathrm{q}}_k)=\sqrt{s}$.
This regularization procedure is applied only when the total energy
$\sqrt{s}$ exceeds the threshold energy $E_k^{\mathrm{thr}}$ of the
$k$-th channel.

In the previous study~\cite{Clymton:2024fbf}, the transition
amplitudes $\mathcal{T}_{J/\psi N,k}$ were computed by including only
the $t$-channel diagrams in the kernel amplitudes. To describe
$J/\psi$ photoproduction quantitatively, we also need to introduce the
$u$-channel diagrams involving heavy-baryon exchange, which provide
significant contributions to the transition amplitudes. Since the
$J/\psi N$ elastic scattering amplitude is small and the transition
amplitudes to the $J/\psi N$ channel are governed by heavy-hadron
exchange, the resonance properties remain almost intact despite the
inclusion of the $u$-channel diagrams.
\begin{figure}[ht]
\centering
\includegraphics[scale=0.35]{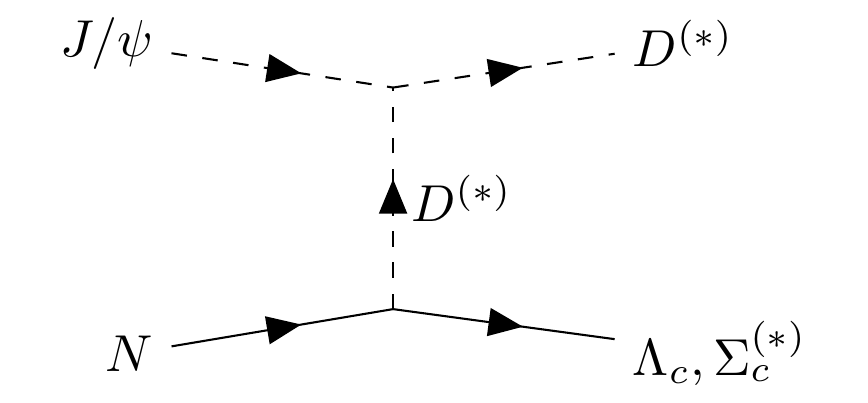}
\includegraphics[scale=0.35]{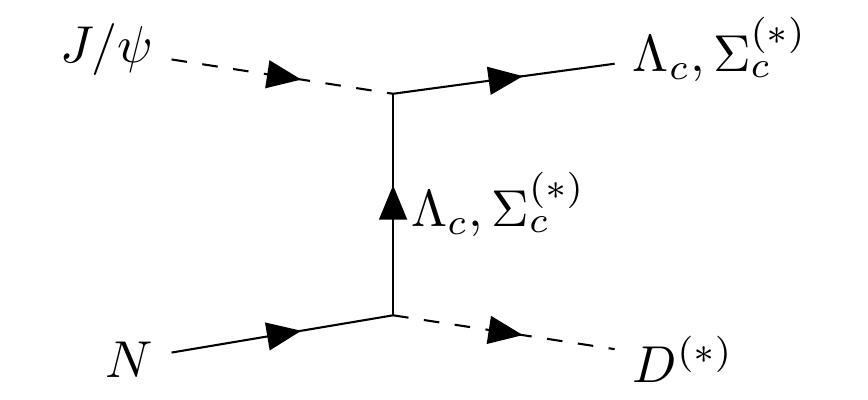}
\caption{The $t$-channel (left panel) and $u$-channel (right panel)
  diagrams.}
\label{fig:2}
\end{figure}
Figure~\ref{fig:2} illustrates the $t$-channel meson-exchange and
$u$-channel baryon-exchange diagrams for the transition to the
$J/\psi N$ channel. The exchanged particles considered in each
reaction are summarized in Table~\ref{tab:1}. A kernel amplitude
associated with a given exchange diagram is written as
\begin{align}
\mathcal{V} =   \mathcal{C}_I\,F^{2}(q^{2})\,\Gamma_{1}\,
  \mathcal{P}(q)\,\Gamma_{2},
\label{eq:6}
\end{align}
where $\mathcal{C}_I$ denotes the isospin factor for each exchanged
particle in a given channel, as listed in Table~\ref{tab:1}. The
vertex functions $\Gamma_{1,2}$ are derived from the following
effective Lagrangians respecting heavy-quark spin symmetry:
\begin{align}
\mathcal{L}_{DDJ/\psi} &= -ig_\psi M_D\sqrt{M_{J}}
  \left(J/\psi^\mu D^\dagger\,\overleftrightarrow{\partial_\mu}D\right),   \cr
\mathcal{L}_{D^*DJ/\psi} &= ig_\psi\sqrt{\frac{M_DM_{D^*}}{M_{J}}}
  \epsilon^{\mu\nu\alpha\beta} \partial_\mu J/\psi_\nu
  \left(D^\dagger\,\overleftrightarrow{\partial_\alpha}D^*_\beta
  +D_{\beta}^{*\dagger}\,\overleftrightarrow{\partial_\alpha}D\right),   \cr
\mathcal{L}_{D^*D^*J/\psi} &= ig_\psi M_{D^*}\sqrt{M_J}\,
  (g^{\mu\nu}g^{\alpha\beta}-g^{\mu\alpha}g^{\nu\beta}+g^{\mu\beta}g^{\nu\alpha})
  \left(J/\psi_\mu
  D_{\nu}^{*\dagger}\,\overleftrightarrow{\partial_\alpha}D^*_\beta\right),   \cr
\mathcal{L}_{DN\Lambda_c} &= -ig_{I\bar{3}}\sqrt{M_{D}}\,
  \bar{N}\gamma_5\Lambda_c\bar{D} +\mathrm{h.c.},   \cr
\mathcal{L}_{D^*N\Lambda_c} &= -ig_{I\bar{3}}\sqrt{M_{D^*}}\,
  \bar{N}\gamma^\mu\Lambda_c\bar{D}^{*}_\mu +\mathrm{h.c.},   \cr
\mathcal{L}_{DN\Sigma_c} &= ig_{I6}\sqrt{\frac{3M_{D}}{2}}\,
  \bar{N}\gamma_5\mathbf{\tau}\cdot\mathbf{\Sigma}_c\bar{D} +\mathrm{h.c.},   \cr
\mathcal{L}_{D^*N\Sigma_c} &= -ig_{I6}\sqrt{\frac{M_{D^*}}{6}}\,
  \bar{N}\gamma^\mu\mathbf{\tau}\cdot\mathbf{\Sigma}_c\bar{D}^*_\mu +\mathrm{h.c.},   \cr
\mathcal{L}_{D^*N\Sigma_c^*} &= ig_{I6}\sqrt{2M_{D^*}}\,
  \bar{N}\gamma_5\mathbf{\tau}\cdot\mathbf{\Sigma}_c^{*\mu}\bar{D}^*_\mu +\mathrm{h.c.},   \cr
\mathcal{L}_{J/\psi\Lambda_c\Lambda_c} &= -g_{J\bar{3}}\,
  \bar{\Lambda}_c\gamma^\mu \Lambda_cJ/\psi_\mu,   \cr
\mathcal{L}_{J/\psi\Sigma_c\Sigma_c} &= -g_{J6}\,
  \bar{\mathbf{\Sigma}}_c\cdot\gamma^\mu \mathbf{\Sigma}_cJ/\psi_\mu,   \cr
\mathcal{L}_{J/\psi\Sigma_c\Sigma_c^*} &= \frac{2g_{J6}}{\sqrt{3}}\,
  \bar{\mathbf{\Sigma}}_c\cdot\gamma_5
  \mathbf{\Sigma}_c^{*\mu}J/\psi_\mu +\mathrm{h.c.},   \cr
\mathcal{L}_{J/\psi\Sigma_c^*\Sigma_c^*} &= g_{J6}\,
  \bar{\mathbf{\Sigma}}_{c\mu}^*\cdot\gamma^\nu
  \mathbf{\Sigma}_c^{*\mu}J/\psi_\nu.
\label{eq:7}
\end{align}

\begin{table}[htbp]
  \caption{\label{tab:1}
    Isospin factors ($\mathcal{C}_I$) for the exchange diagrams contributing
    to the transitions from the $J/\psi N$ channel to each open-charm
    meson-baryon channel, together with the corresponding exchanged
    particles.}
  \begin{ruledtabular}
  \centering\begin{tabular}{lcr}
   Reactions & Exch. & $\mathcal{C}_I$
   \\\hline
     $J/\psi N\to\bar{D}\Lambda_c$
     & $\bar{D}$, $\bar{D}^*$, $\Lambda_c$ & $1$ \\
     $J/\psi N\to\bar{D}^*\Lambda_c$
     & $\bar{D}$, $\bar{D}^*$, $\Lambda_c$ & $1$ \\
     $J/\psi N\to\bar{D}\Sigma_c$
     & $\bar{D}$, $\bar{D}^*$, $\Sigma_c$ & $\sqrt{3}$ \\
     $J/\psi N\to\bar{D}\Sigma_c^*$
     & $\bar{D}^*$, $\Sigma_c$ & $\sqrt{3}$ \\
     $J/\psi N\to\bar{D}^*\Sigma_c$
     & $\bar{D}$, $\bar{D}^*$, $\Sigma_c$, $\Sigma_c^*$ & $\sqrt{3}$ \\
     $J/\psi N\to\bar{D}^*\Sigma_c^*$
     & $\bar{D}^*$, $\Sigma_c$, $\Sigma_c^*$ & $\sqrt{3}$ \\
  \end{tabular}
  \end{ruledtabular}
\end{table}

Since no reliable theoretical information is available for the
relevant coupling constants, we estimate them using SU(4) symmetry
relations, even though this symmetry is strongly broken. To quantify
the uncertainty arising from SU(4) symmetry, we introduce a common
prefactor $a$ by which all couplings derived from the SU(4) relations
are multiplied. The couplings $g_{I\bar{3}}$ and $g_{I6}$ are then
estimated as 
\begin{align}
g_{I\bar{3}}\sqrt{M_D} &= -\frac{3\sqrt{3}}{5}\,ag_{\pi NN},   \cr
g_{I6}\sqrt{\frac{3M_D}{2}} &= -\frac{ag_{\pi NN}}{5}.
\label{eq:8}
\end{align}
Similarly, $g_{J\bar{3}}$ and $g_{J6}$ are obtained from $g_{\rho NN}$
via the SU(4) relation:
\begin{align}
g_{J\bar{3}} = g_{J6} = \sqrt{2}\,ag_{\rho NN}.
\label{eq:9}
\end{align}
In the present work, we adopt the coupling constants from the Nijmegen
potential~\cite{Stoks:1999bz,Rijken:1998yy}, namely $g_{\pi NN}=13.2$
and $g_{\rho NN}=2.97$.

For the mesonic sector, we follow the same procedure and estimate the
$g_{\psi}$ coupling as
\begin{align}
g_{\psi}\,M_D\sqrt{M_J} = \frac{\sqrt{2}}{2}\,ag_{\pi\pi\rho}.
\label{eq:10}
\end{align}
The value $g_{\pi\pi\rho}=5.97$ is taken from our previous
work~\cite{Clymton:2022jmv}. We find that the parameter $a$ controls
the overall magnitude of the total $J/\psi$ photoproduction cross
section without significantly affecting the hidden-charm pentaquark
resonances. A value of $a=0.47$ provides the best description of the
experimental data, whereas $a=1$ yields a cross section roughly an
order of magnitude larger. Thus, the parameter $a$ is the main source
for the uncertainty of the present work. 

The propagators for pseudoscalar and vector meson exchange are given by~\cite{Kim:2024mqx}
\begin{align}
  \mathcal{P}(q,M) &= \frac{1}{q^2-M^2},   \cr
  \mathcal{P}_{\mu\nu}(q,M) &= \frac{1}{q^2-M^2}
  \left(-g_{\mu\nu}+\frac{q_\mu q_\nu}{M^2}\right),
\label{eq:11}
\end{align}
whereas those for spin-$\frac{1}{2}$ and spin-$\frac{3}{2}$ baryon
exchange are expressed as 
\begin{align}
  \mathcal{P}(q,M) &= \frac{\slashed{q}+M}{q^2-M^2},   \cr
  \mathcal{P}_{\mu\nu}(q,M) &= \frac{\slashed{q}+M}{q^2-M^2}
  \left(-g_{\mu\nu}+\frac{1}{3}\gamma_\mu\gamma_\nu
  +\frac{1}{3M}(\gamma_\mu q_\nu-\gamma_\nu q_\mu)
  +\frac{2}{3M^2}q_\mu q_\nu\right).
\label{eq:12}
\end{align}
Here, $q$ denotes the exchanged four-momentum and $M$ the mass of the
exchanged particle.

Since hadrons have finite sizes, a form factor is introduced at each
vertex. This form factor is also necessary to ensure the unitarity of
the transition amplitudes. We use the following
form~\cite{Kim:1994ce}:
\begin{align}
F_{n}(q^{2}) =
  \left(\frac{n\Lambda^{2}-M^{2}}{n\Lambda^{2}-q^{2}}\right)^{n},
  \label{eq:13}
\end{align}
where the positive integer $n$ is chosen such that the power of
momentum in the vertex function $\Gamma$ is regularized. As
$n\to\infty$, the form factor in Eq.~\eqref{eq:13} approaches a
Gaussian form. Note that the energy dependence in Eq.~\eqref{eq:13}
must be removed, as it leads to unphysical behavior in the kernel
amplitude. Although the values of the cutoff masses $\Lambda$ in
Eq.~\eqref{eq:13} cannot be determined directly from experiment or
theory, the associated uncertainties can be reduced by exploiting the
physical properties of the hadrons involved. Since heavy hadrons are
more compact than lighter ones~\cite{Kim:2018nqf,Kim:2021xpp}, higher
cutoff masses are appropriate for them, as the cutoff mass is inversely
proportional to the size of the corresponding hadron. We therefore
introduce the reduced cutoff mass $\Lambda_0 := \Lambda - M$. In
Ref.~\cite{Cheng:2004ru}, the cutoff mass was shown to be related to
$\Lambda_{\mathrm{QCD}}$ via $\Lambda = M + \eta
\Lambda_{\mathrm{QCD}}$, where $\eta$ is of order unity. This approach
has been successfully applied in previous
studies~\cite{Clymton:2022jmv,Clymton:2023txd,Kim:2023htt,Clymton:2024pql, 
Kim:2025ado,Clymton:2024fbf,Clymton:2025hez,Clymton:2025zer}. In this
work, all values of the reduced cutoff mass are fixed at
$\Lambda_0 = 600$~MeV. We emphasize that we do not change the values
of $\Lambda_0$ for fitting. 

Having revised the kernel amplitudes for the transition to the
$J/\psi N$ channel, we recalculate the transition amplitudes
$\mathcal{T}_{J/\psi N,k}$ within the same framework as in
Ref.~\cite{Clymton:2024fbf}. The resulting $J/\psi N$ transition
amplitudes are discussed in detail below. Throughout the calculation,
isospin symmetry is assumed for the hadron masses, and this
approximation is retained exclusively for the masses in the remainder
of the calculation.
\begin{figure*}[ht]
\centering
\includegraphics[scale=0.45]{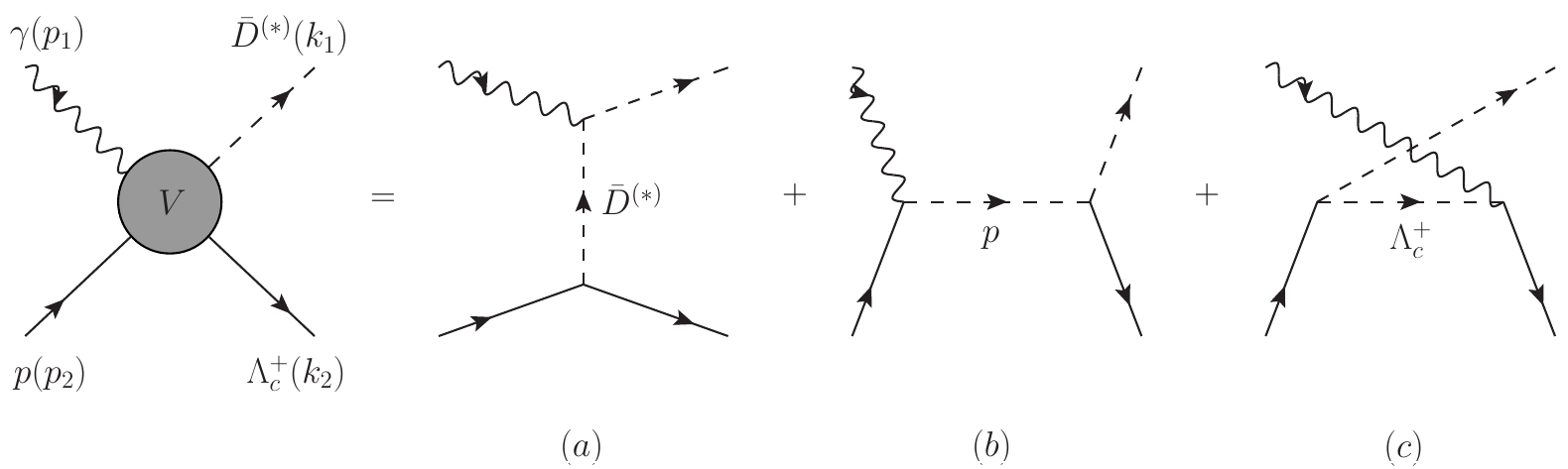}
\caption{Tree-level diagrams for the subprocess $\gamma p \to
\bar{D}^{(*)} \Lambda_c\,(\Sigma_c)$ contributing to the rescattering
amplitude: (a) $D^{(*)}$-meson exchange in the $t$ channel, (b) nucleon
exchange in the $s$ channel, and (c) $\Lambda_c$ ($\Sigma_c$) baryon
exchange in the $u$ channel.}
\label{fig:3}
\end{figure*}
The kernel amplitudes $V_{k,\gamma p}$ in Eq.~\eqref{eq:1} receive
contributions from the three tree-level diagrams depicted in
Fig.~\ref{fig:3}: (a) $D^{(*)}$ exchange in the $t$ channel, (b)
nucleon exchange in the $s$ channel, and (c) $\Lambda_c$ or $\Sigma_c$
exchange in the $u$ channel. All possible exchange diagrams for each
reaction channel are listed in Table~\ref{tab:2}.
\begin{table}[htbp]
  \caption{\label{tab:2}
    Exchange diagrams contributing to each charmed meson--baryon
    photoproduction channel off the proton.}
  \begin{ruledtabular}
  \centering\begin{tabular}{lr}
   Reaction & Exchanged particles
   \\\hline
     $\gamma p\to\bar{D}^0\Lambda_c^+$
     & $p$,\; $\bar{D}^{*0}$,\; $\Lambda_c^+$ \\
     $\gamma p\to\bar{D}^{*0}\Lambda_c^+$
     & $p$,\; $\bar{D}^{0}$,\; $\Lambda_c^+$ \\
     $\gamma p\to\bar{D}^{0}\Sigma_c^+$
     & $p$,\; $\bar{D}^{*0}$,\; $\Sigma_c^+$ \\
     $\gamma p\to D^{-}\Sigma_c^{++}$
     & $p$,\; $D^{*-}$,\; $\bar{D}^{*0}$,\; $\Sigma_c^{++}$ \\
     $\gamma p\to\bar{D}^{*0}\Sigma_c^+$
     & $p$,\; $\bar{D}^{0}$,\; $\Sigma_c^+$ \\
     $\gamma p\to D^{*-}\Sigma_c^{++}$
     & $p$,\; $D^{*-}$,\; $\bar{D}^{0}$,\; $\Sigma_c^{++}$
  \end{tabular}
  \end{ruledtabular}
\end{table}
The corresponding vertex functions are derived from the effective
Lagrangians in Eq.~\eqref{eq:7} together with the following
electromagnetic interaction Lagrangians:
\begin{align}
\mathcal{L}_{\gamma DD} &=
  -ie_{\bar{D}}A_\mu(\bar{D}\,\partial^\mu D-\partial^\mu\bar{D}\,D),   \cr
\mathcal{L}_{\gamma D D^*} &=
  ig_{\gamma D D^*} \epsilon^{\mu\nu\alpha\beta} \partial_\mu A_\nu
  \left(\bar D \,\overleftrightarrow{\partial_\alpha} D^{*}_\beta
  + \bar D^{*}_\beta \,\overleftrightarrow{\partial_\alpha} D\right),   \cr
\mathcal{L}_{\gamma D^*D^*} &=
  ie_{\bar{D}^*} (g^{\mu\nu}g^{\alpha\beta}-g^{\mu\alpha}g^{\nu\beta}
  +g^{\mu\beta}g^{\nu\alpha})A_\mu
  \left(\bar{D}^*_\nu\,\partial_\alpha D^*_\beta
  -\partial_\alpha\bar{D}^*_\nu D^*_\beta\right),   \cr
\mathcal{L}_{\gamma NN} &=
  -e_N\,\bar{N}\left[\gamma^\mu
  -\frac{\kappa_N}{2M_N}\sigma^{\mu\nu}\partial_\nu\right]A_\mu N,   \cr
\mathcal{L}_{\gamma \Lambda_c \Lambda_c} &=
  -e_{\Lambda_c}\,\bar{\Lambda}_c\left[\gamma^\mu
  -\frac{\kappa_{\Lambda_c}}{2M_{\Lambda_c}}\sigma^{\mu\nu}
  \partial_\nu\right]A_\mu\Lambda_c,   \cr
\mathcal{L}_{\gamma \Sigma_c \Sigma_c} &=
  -e_{\Sigma_c}\,\bar{\Sigma}_c\left[\gamma^\mu
  -\frac{\kappa_{\Sigma_c}}{2M_{\Sigma_c}}\sigma^{\mu\nu}
  \partial_\nu\right]A_\mu\Sigma_c.
\label{eq:14}
\end{align}
The relevant coupling constants are determined as
\begin{align}
&g_{\gamma D^0 D^{*0}} = 0.30\;\mathrm{GeV}^{-1},\quad
  g_{\gamma D^- D^{*-}} = 0.141\;\mathrm{GeV}^{-1},   \cr
&\kappa_p = 1.79,\quad
  \kappa_{\Lambda_c} = -0.025,\quad
  \kappa_{\Sigma_c^{+}} = 0.41,\quad
  \kappa_{\Sigma_c^{++}} = 4.64,
\label{eq:15}
\end{align}
where $g_{\gamma DD^{*}}$ is extracted from the branching ratio
$\mathrm{Br}(D^{*}\to D\gamma)$ and the total decay width
$\Gamma_{D^{*}}$~\cite{PDG:2024cfk, Rosner:2013sha} via the relation
\begin{align}
\Gamma_{D^{*} \to D \gamma} =
\frac{q_\gamma^3}{3\pi}\,(g_{\gamma D D^{*}})^2,
\label{eq:16}
\end{align}
with $q_\gamma = (M_{D^{*}}^2-M_{D}^2)/(2M_{D^{*}})$. The anomalous
magnetic moment of the proton is taken from Ref.~\cite{PDG:2024cfk}.
The anomalous magnetic moment of $Y_c$ (denoting either $\Lambda_c$ or
$\Sigma_c$) is related to its magnetic moment $\mu_{Y_c}$~\cite{Fomin:2019wuw} by
\begin{align}
\mu_{Y_c} = \frac{e_{Y_c}+e\kappa_{Y_c}}{2M_{Y_c}}.
\label{eq:17}
\end{align}

The transition amplitudes are formulated to ensure gauge invariance, 
which is why we include the proton-pole diagram despite its small
contribution. Notably, the kernel amplitudes are governed by distinct
coupling factors: $V_{\bar{D}^{(*)}\Lambda_c,\gamma p}$ is proportional
to $g_{\bar{D}^{(*)}N\Lambda_c}$, while
$V_{\bar{D}^{(*)}\Sigma_c,\gamma p}$ is proportional to 
$g_{\bar{D}^{(*)}N\Sigma_c}$. This distinction is a crucial feature, as
it explains the relative suppression of the $\bar{D}^{(*)}\Sigma_c$
rescattering contribution.

In the present study, we exclude the $\bar{D}^{(*)}\Sigma_c^*$
intermediate states from the rescattering calculation for two primary
reasons. Firstly, the coupling constant
$g_{\bar{D}^{(*)}N\Sigma_c^{(*)}}$ is approximately five times smaller
than $g_{\bar{D}^{(*)}N\Lambda_c}$, so that contributions from the
$\bar{D}^{(*)}\Sigma_c^{(*)}$ intermediate states are significantly
suppressed relative to those from $\bar{D}^{(*)}\Lambda_c$. Secondly,
constructing gauge-invariant amplitudes that involve a
spin-$\frac{3}{2}$ exchange propagator presents considerable technical
difficulties. For these reasons, the contributions from the
$\bar{D}^{(*)}\Sigma_c^*$ channels are omitted.

Having formulated the kernel amplitude $V_{k,\gamma p}$, we proceed to
evaluate the rescattering amplitude. Since the transition amplitude for
the $J/\psi N$ channel was computed in the isospin basis, the
corresponding amplitudes for definite charge states are related by
\begin{align}
\mathcal{T}_{J/\psi p, \bar{D}^0\Lambda_c^+} = \mathcal{T}_{J/\psi N,
  \bar{D}\Lambda_c}, 
\hspace{0.5cm}
\mathcal{T}_{J/\psi p, \bar{D}^0\Sigma_c^+} =
  \frac{1}{\sqrt{3}}\,\mathcal{T}_{J/\psi N, \bar{D}\Sigma_c}, 
\hspace{0.5cm}
\mathcal{T}_{J/\psi p, D^-\Sigma_c^{++}} =
  -\sqrt{\frac{2}{3}}\,\mathcal{T}_{J/\psi N, \bar{D}\Sigma_c}. 
\end{align}

For the background contribution, we employ the pomeron-exchange
amplitude of
Refs.~\cite{Donnachie:1984xq,Donnachie:1985iz,Donnachie:1987pu}, which 
has been shown to describe successfully the total cross-section data
for diffractive vector-meson photoproduction in the high-energy regime
($E_\gamma \geqslant 10$~GeV). The total cross section for $J/\psi$
photoproduction off the proton is then obtained by combining the
rescattering and pomeron-exchange amplitudes.

\section{Numerical Results} \label{sec:3}
The null result for the hidden-charm pentaquark signal in
$J/\psi$ photoproduction has so far remained unexplained. Previous
attempts employed the meson-exchange picture combined with pomeron
contributions, which reproduce the data reasonably well, especially
when the two $\bar{D}^{(*)}\Lambda_c$ channels are included through a
rescattering process that generates threshold effects at their
respective thresholds~\cite{Du:2020bqj}. These studies, however, do
not include the $P_{c\bar{c}}$ states and therefore cannot explain how the
signal disappears. In the present work, we investigate $J/\psi$
photoproduction off the proton using the transition amplitudes
obtained in Ref.~\cite{Clymton:2024fbf}, which contain dynamically
generated hidden-charm pentaquark states consistent with the
candidates observed in $\Lambda_b$ decays by the LHCb Collaboration. 

\begin{figure}[ht]
\centering
\includegraphics[scale=0.56]{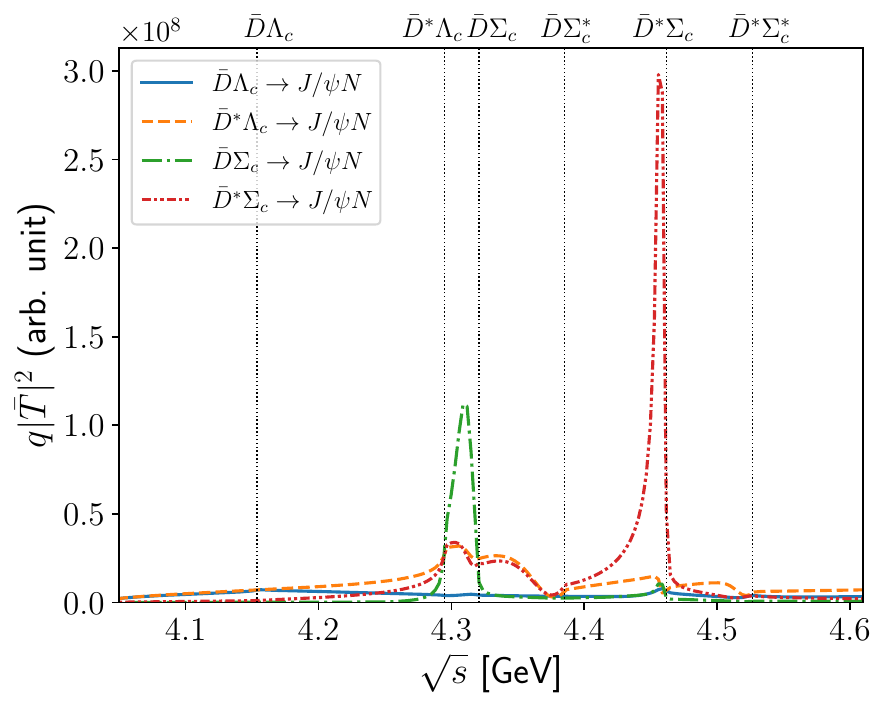}
\caption{Spin-averaged squared transition amplitude to the $J/\psi N$
  channel, multiplied by the final-state center-of-mass momentum,
  $q\,|\bar{T}|^2$, as a function of the total energy $\sqrt{s}$. The
  vertical dashed lines indicate the relevant meson--baryon
  thresholds.}
\label{fig:4}
\end{figure}
Before evaluating the $J/\psi$ photoproduction results, we discuss the
$J/\psi N$ transition amplitudes. Figure~\ref{fig:4} shows the
distributions of the transition amplitudes from the
$\bar{D}\Lambda_c$, $\bar{D}^*\Lambda_c$, $\bar{D}\Sigma_c$, and
$\bar{D}^*\Sigma_c$ channels to the $J/\psi N$ state,
respectively. We restrict ourselves to these four transitions,
since the $\bar{D}^{(*)}\Sigma_c^*$ intermediate states were excluded
from the rescattering contribution to the $J/\psi$ photoproduction
amplitudes, as justified in the previous section. The $P_{c\bar{c}}$
states appear prominently in the $\bar{D}\Sigma_c \to J/\psi N$ and 
$\bar{D}^*\Sigma_c \to J/\psi N$ transitions, with magnitudes
exceeding those of the $\bar{D}^{(*)}\Lambda_c$ channels. This
qualitatively implies that the resonance structure should be clearly
observable unless the $\bar{D}^{(*)}\Sigma_c$ channels are
significantly suppressed. 

\begin{figure}[ht]
\centering
\includegraphics[scale=0.56]{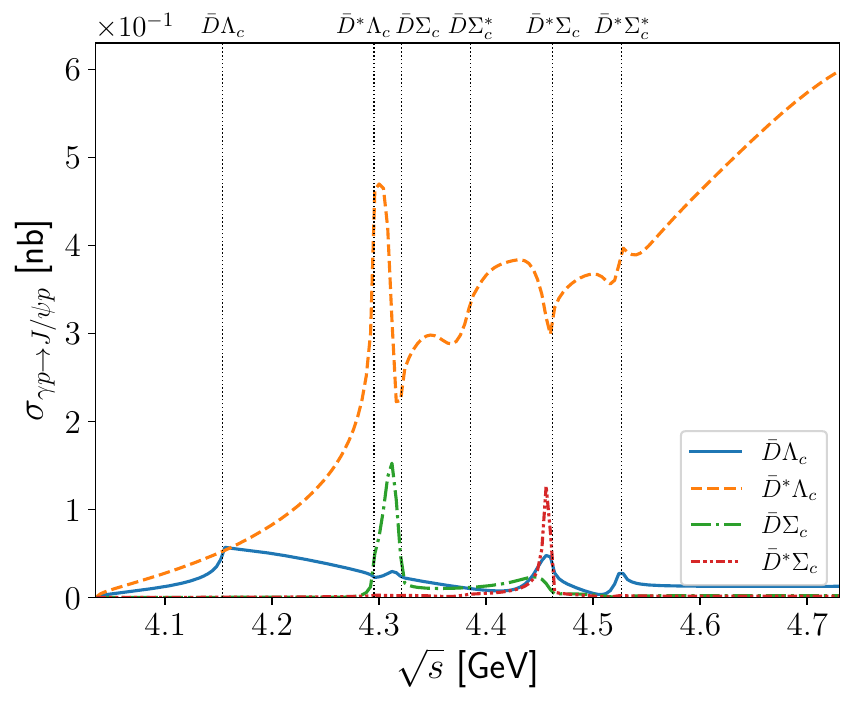}
\caption{Total cross section for the $\gamma p\to J/\psi\,p$ reaction
  from the rescattering contribution, evaluated separately for each
  intermediate state as a function of the total energy $\sqrt{s}$. The
  vertical dashed lines indicate the relevant meson--baryon
  thresholds.}
\label{fig:5}
\end{figure}
As is evident from the structure of the kernel amplitude
$V_{k,\gamma p}$, the rescattering contributions from the
$\bar{D}^{(*)}\Sigma_c$ intermediate states are suppressed. To
demonstrate this explicitly, Fig.~\ref{fig:5} compares the total cross
section from the rescattering part, evaluated for each intermediate
channel. The contributions from the $\bar{D}\Sigma_c$ and
$\bar{D}^*\Sigma_c$ channels are approximately one order of magnitude
smaller than those from the $\bar{D}\Lambda_c$ and
$\bar{D}^*\Lambda_c$ channels. This illustrates why the
$P_{c\bar{c}}$ states are suppressed in the photoproduction process:
the $\bar{D}^{(*)}\Sigma_c$ channels couple only weakly to $\gamma N$.

\begin{figure}[ht] 
\centering
\includegraphics[scale=0.55]{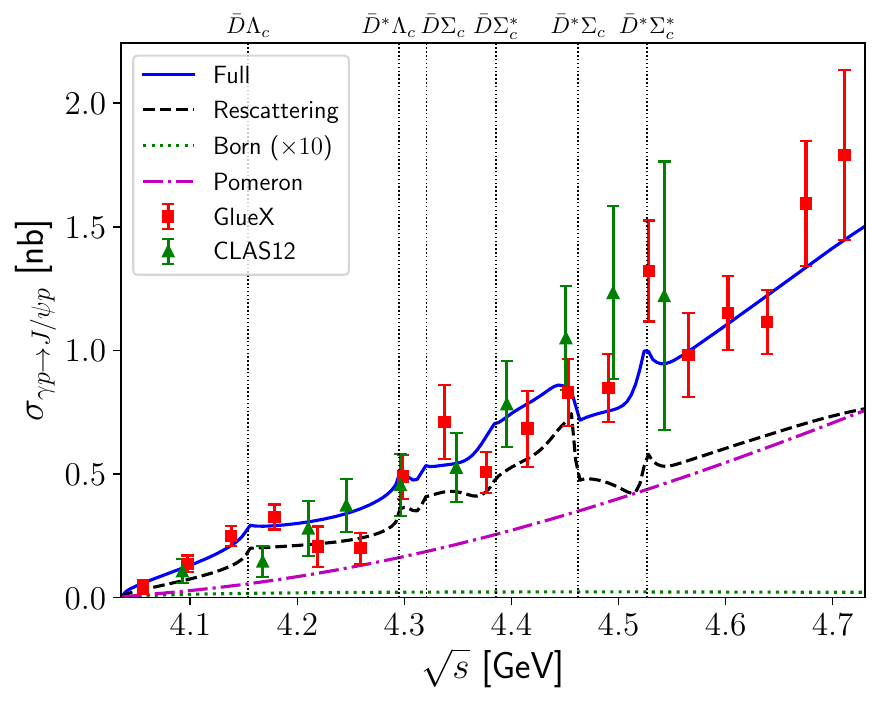}
\caption{Total cross section as a function of the total energy.
  Experimental data are taken from Ref.~\cite{GlueX:2023pev} (squares)
  and Ref.~\cite{CLAS:2026lls} (triangles). The error bars represent
  the statistical and systematic uncertainties added in quadrature.
  The vertical dashed lines indicate the relevant meson--baryon
  thresholds.}
\label{fig:6}
\end{figure}

Furthermore, a close inspection of the small peak structure arising
from $\bar{D}^*\Lambda_c$ rescattering reveals a weak signal of
$P_{c\bar{c}}(4312)$. However, interference from the
$\bar{D}\Sigma_c$ channel acts destructively, particularly in the
$P_{c\bar{c}}(4312)$ peak region. Consequently, no visible peak
remains in the total rescattering contribution to $J/\psi$
photoproduction, as indicated by the dashed line in
Fig.~\ref{fig:6}.

Having discussed the rescattering mechanism, we now examine the total
cross section for $J/\psi$ photoproduction off the proton as a
function of energy, comparing the results with the experimental data
in Fig.~\ref{fig:6}. The present model is constructed primarily from
the pomeron-exchange and rescattering contributions, since the Born
term has a nearly negligible effect. The results indicate that the
sizable cross section for $J/\psi$ production originates from the
rescattering effect, which facilitates processes that would otherwise
be OZI-suppressed. The model reproduces the experimental data from
both the GlueX and CLAS12 Collaborations remarkably well. As discussed
above, the present model supports the nonexistence of hidden-charm
pentaquarks in $J/\psi$ photoproduction, owing to the suppression of
the $\bar{D}^{(*)}\Sigma_c$ rescattering contribution, as mentioned
already. 

While the GlueX experiment shows no signal for the hidden-charm
pentaquark states, it exhibits cusp structures at the
$\bar{D}^*\Lambda_c$, $\bar{D}\Sigma_c$, and $\bar{D}^*\Sigma_c^*$
thresholds~\cite{GlueX:2023pev, Du:2020bqj}. These cusp structures
arise mainly from the coupled-channel rescattering effects. In
particular, we find that they are enhanced by the inclusion of the
baryon-exchange diagrams, which were not considered in our previous
work~\cite{Clymton:2024fbf}. Baryon exchange is therefore essential in
describing the cusp structures. As shown in Fig.~\ref{fig:6}, the
present results for the cusp structures are consistent with the
experimental data.
\begin{figure}[ht] 
\centering
\includegraphics[scale=0.55]{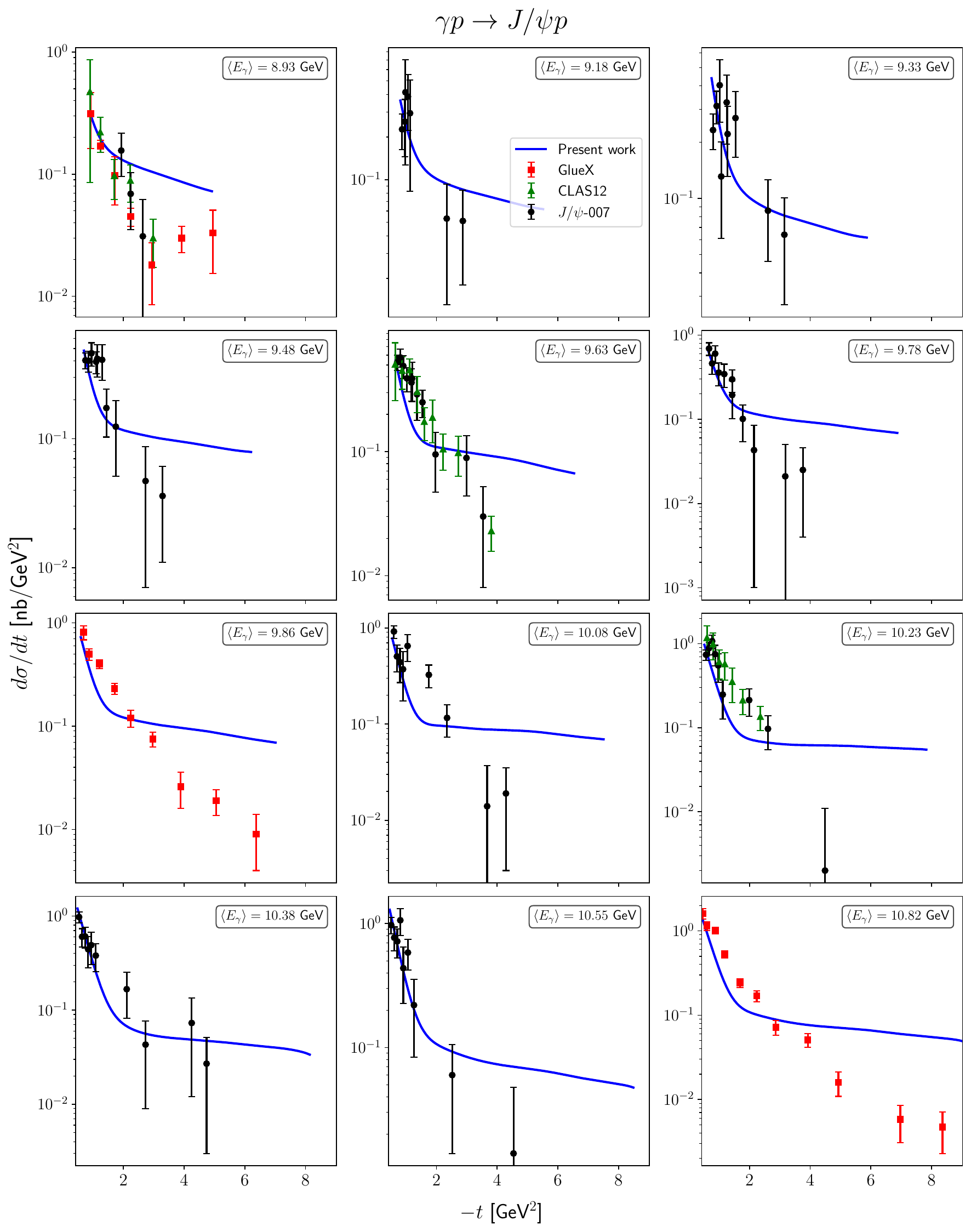}
\caption{Differential cross section as a function of the exchange
  momentum transfer $-t$. Experimental data are taken from
  Ref.~\cite{GlueX:2023pev} (squares), Ref.~\cite{CLAS:2026lls}
  (triangles), and Ref.~\cite{Duran:2022xag} (circles). The error bars
  represent the statistical and systematic uncertainties added in
  quadrature.}
\label{fig:7}
\end{figure}
In Fig.~\ref{fig:7}, we present the numerical results for the
differential cross section $d\sigma/dt$ as a function of $-t$ in the
range of the average photon energy $8.93\,\mathrm{GeV} \le \langle
E_\gamma \rangle \le 10.82\,\mathrm{GeV}$. As expected, the data are
described very well in the small $-t$ region. As $-t$ increases, the
present results fall off more slowly than the data; however, the
experimental uncertainties also grow with $-t$.
\section{Summary and conclusions} \label{sec:4}
In this work, we have investigated $J/\psi$ photoproduction off the
proton, $\gamma p \to J/\psi\,p$, to clarify why no hidden-charm
pentaquark signals appear in the GlueX and CLAS12 experiments. The
present framework combines a pomeron-exchange
background with a coupled-channel rescattering mechanism.
The transition amplitudes $\mathcal{T}_{J/\psi N, k}$ in the
rescattering kernel were taken from our coupled-channel analysis,
which dynamically generates hidden-charm
pentaquark states consistent with the $P_{c\bar{c}}$ observed by the
LHCb Collaboration. We have revised the kernel amplitudes for the
transition to the $J/\psi N$ 
channel so as to include both $t$-channel heavy-meson exchange and
$u$-channel heavy-baryon exchange, derived from effective
Lagrangians. The coupling constants were 
estimated from SU(4) symmetry relations, and the resulting uncertainty
was absorbed into a single common prefactor $a$, which controls the
overall magnitude of the cross section. The reduced cutoff mass was
kept at $\Lambda_0 = 600$~MeV
throughout, and the value $a = 0.47$ was the only quantity adjusted to
the data.

In this study, we provide a theoretical mechanism
for the suppression of the hidden-charm pentaquark signal in $J/\psi$
photoproduction. The rescattering
contributions from the $\bar{D}^{(*)}\Sigma_c$ intermediate channels
are about one order of magnitude smaller than those from
$\bar{D}^{(*)}\Lambda_c$. This difference originates from the coupling
constants in the photoproduction kernel:
$V_{\bar{D}^{(*)}\Sigma_c,\gamma p}$ is governed by
$g_{\bar{D}^{(*)}N\Sigma_c}$, which is about one-fifth of
$g_{\bar{D}^{(*)}N\Lambda_c}$. Since the $P_{c\bar{c}}$ resonances
couple to the $J/\psi N$ channel predominantly through the
$\bar{D}^{(*)}\Sigma_c$ intermediate states, the strong suppression of
these channels prevents the pentaquark signal from appearing in
$J/\psi$ photoproduction off the proton. The weak signal of
$P_{c\bar{c}}(4312)$ that would otherwise emerge through
$\bar{D}^*\Lambda_c$ rescattering is further diminished by destructive
interference from the $\bar{D}\Sigma_c$ channel.

While the pentaquark signal is suppressed, the GlueX data exhibit cusp
structures at the meson-baryon thresholds, which arise mainly from the
coupled-channel rescattering effects. We have found that these cusps
are enhanced by the inclusion of the $u$-channel baryon-exchange
diagrams, not considered in our previous work, so that baryon
exchange is essential for describing them. With these
ingredients, the total cross section agrees well with the
GlueX and CLAS12 data over a
wide energy range, and the differential cross section as a function of
$-t$ is well reproduced in the small $-t$ region. These results
confirm that the sizable $J/\psi$ photoproduction cross section is
driven primarily by the rescattering effect through open-charm
meson-baryon loops.

In conclusion, we have provided a natural explanation for the
nonexistence of hidden-charm pentaquark signals in $J/\psi$
photoproduction: the $\bar{D}^{(*)}\Sigma_c$ rescattering channels,
which are indispensable for the formation of the $P_{c\bar{c}}$ states,
couple only weakly to the $\gamma N$ initial state because
$g_{\bar{D}^{(*)}N\Sigma_c}$ is much smaller than
$g_{\bar{D}^{(*)}N\Lambda_c}$. This contrasts with $\Lambda_b$ decays,
where the $\bar{D}\Sigma_c$ channel acts as a filter that selects the
pentaquark states. The present results therefore
suggest that the null result from GlueX and CLAS12 need not be in
conflict with the pentaquark signals observed by the LHCb
Collaboration, and that the absence of a pentaquark signal in
photoproduction is not by itself evidence against the existence of
hidden-charm pentaquark baryons.

\section*{Acknowledgments}
The present work was supported by the Young Scientist Training (YST)
Program at the Asia Pacific Center for Theoretical Physics (APCTP)
through the Science and Technology Promotion Fund and Lottery Fund of
the Korean Government and also by the Korean Local Governments –
Gyeongsangbuk-do Province and Pohang City (SC), the Basic Science
Research Program through the National Research Foundation of Korea 
(NRF), Grants No. RS-2021-NR060129 (SHK) and No. RS-2025-00513982 
(HChK). 

\bibliography{Jpsi}

\end{document}